\documentclass[conference]{IEEEtran}
\IEEEoverridecommandlockouts

\usepackage{cite}
\usepackage{amsmath,amssymb,amsfonts}
\usepackage{graphicx}
\usepackage{textcomp}
\usepackage{xcolor}
 \usepackage{booktabs}
\usepackage{listings}
\usepackage{algorithm}
\usepackage{algpseudocode}
\usepackage{hyperref}
\def\BibTeX{{\rm B\kern-.05em{\sc i\kern-.025em b}\kern-.08em
    T\kern-.1667em\lower.7ex\hbox{E}\kern-.125emX}}
\begin{document}

\title{Whitespaces Don’t Lie: Feature-Driven and Embedding-Based Approaches for Detecting Machine-Generated Code}

\author{
\IEEEauthorblockN{Syed Mehedi Hasan Nirob}
\IEEEauthorblockA{\textit{Computer Science and Engineering} \\
\textit{Shahjalal University of
Science and Technology}\\
Sylhet-3114, Bangladesh \\
smh.nirob@gmail.com}
\and
\IEEEauthorblockN{Shamim Ehsan}
\IEEEauthorblockA{University of Texas at El Paso\\
El Paso, TX, 79968, USA \\
sehsan@miners.utep.edu}
\and
\IEEEauthorblockN{Moqsadur Rahman}
\IEEEauthorblockA{Assistant Professor \\
Shahjalal University of Science and Technology\\
Sylhet-3114, Bangladesh \\
moqsad-cse@sust.edu}
\and
\IEEEauthorblockN{Summit Haque}
\IEEEauthorblockA{Assistant Professor \\
Shahjalal University of Science and Technology\\
Sylhet-3114, Bangladesh \\
summit-cse@sust.edu}
}


\maketitle

\begin{abstract}
Large language models (LLMs) have made it remarkably easy to synthesize plausible source code from natural language prompts. While this accelerates software development and supports learning, it also raises \emph{new risks} for academic integrity, authorship attribution, and responsible AI use. This paper investigates the problem of distinguishing \emph{human-written} from \emph{machine-generated} code by comparing two complementary approaches: feature based detectors built from lightweight, interpretable stylometric and structural properties of code, and embedding-based detectors leveraging pretrained code encoders. Using a recent large-scale benchmark dataset of 600k human written and AI-generated code samples, we find that feature-based models achieve strong performance (ROC--AUC 0.995, PR--AUC 0.995, F1 0.971), while embedding-based models with CodeBERT embeddings are also very competitive (ROC--AUC 0.994, PR--AUC 0.994, F1 0.965). Analysis shows that features tied to indentation and whitespace provide particularly discriminative cues, whereas embeddings capture deeper semantic patterns and yield slightly higher precision. These findings underscore the trade-offs between interpretability and generalization, offering practical guidance for deploying robust code-origin detection in academic and industrial contexts.
\end{abstract}

\begin{IEEEkeywords}
Machine-generated code detection, Human vs. AI classification, Handcrafted features, Code embeddings
\end{IEEEkeywords}

\section{Introduction}
The rapid progress of large language models (LLMs) has fundamentally changed the landscape of software development and programming education. Systems such as ChatGPT~\cite{bChatGPT} and GitHub Copilot\cite{bCopilot} can now generate high-quality source code from natural language instructions, complete assignments, and synthesize entire solutions that pass unit tests. For developers, this accelerates prototyping and boosts productivity. For students, it lowers entry barriers to programming and provides personalized assistance. Yet these benefits come with new and urgent risks. In educational settings, students may present AI-generated solutions as original work, undermining fairness and assessment validity. In research and professional contexts, uncertainty about code provenance complicates accountability, licensing, and long-term maintainability. These concerns highlight the urgent need for robust and transparent methods to determine whether a given piece of code was written by a human or generated by a machine.

While the problem of detecting machine-generated content has been studied extensively for natural language text, directly applying these techniques to source code is nontrivial. Code is not only more structured but also more constrained, with formal syntax and strong stylistic signals in whitespace usage, identifier naming, and control-flow structures. At the same time, pretrained models of code can capture deeper semantic regularities, raising the question of whether handcrafted features or learned embeddings provide stronger evidence of authorship. Institutions and practitioners who must operationalize detection therefore face a practical trade-off: should they rely on lightweight, interpretable stylometric detectors, or on resource-intensive embedding-based classifiers?

This question is not merely academic. Universities are increasingly under pressure to ensure assessment integrity in programming courses, plagiarism-detection tools are beginning to integrate AI-origin checks, and industry is exploring safeguards for verifying the provenance of open-source contributions. In each of these contexts, both accuracy and transparency matter: a detector that is overly aggressive risks mislabeling genuine student work as AI-generated, while a detector that is too lenient may fail to flag problematic reliance on generative tools. A systematic and carefully benchmarked comparison of alternative detection pipelines is therefore critical for guiding real-world adoption.

In this work, we address this gap by conducting a systematic approach to machine-generated code detection with feature-based and embedding-based approaches. The feature-based approach extracts interpretable metrics such as whitespace ratios, identifier stylometry, and structural complexity, which may reflect human habits and inconsistencies. The embedding-based approach leverages semantic representations from a pretrained transformer model of code, offering the potential to generalize across styles and unseen generators, but at higher computational cost and reduced interpretability. By placing these two families of methods under a unified evaluation framework, we are able to provide clear evidence of their strengths, weaknesses, and trade-offs.

Our study uses a recent benchmark dataset\cite{semeval}, which provides a large and diverse benchmark of human- and LLM-generated code. We conduct a unified evaluation with consistent preprocessing, and threshold calibration to maximize F1 under imbalance. Findings in our experiment show that feature-based classifiers achieve slightly higher accuracy and recall in validation, driven largely by whitespace- and structure-related cues. Embedding-based classifiers, while marginally lower in raw accuracy, achieve stronger precision and exhibit promising robustness under distribution shift. These complementary strengths suggest that hybrid approaches, which combine interpretable surface features with deep semantic embeddings, may ultimately provide the most reliable detectors for real-world deployment.

The main contributions of this work are as follows:
\begin{itemize}
    \item We design and implement two complementary detection pipelines: one based on handcrafted features of code style and structure, and one based on semantic embeddings from a pretrained transformer model.  
    \item We conduct a systematic comparison of both approaches under a unified evaluation protocol, reporting results across validation and held-out test sets to assess accuracy, interpretability, efficiency, and robustness.
\end{itemize}

\section{Related Work}

The task of distinguishing machine-generated code from human-authored code intersects with several established research areas, including authorship attribution, plagiarism detection, machine-generated text detection, and more recent studies on code representations with pretrained models. In the following, we discuss our contribution within the broader landscape of these domains.

\subsection{Authorship Attribution in Text and Code}
Authorship attribution in natural language has traditionally focused on stylistic features such as vocabulary richness, sentence length, and syntactic patterns~\cite{b1}. These methods demonstrated that individual authors exhibit consistent linguistic fingerprints that can be exploited for attribution tasks. Building on this foundation, researchers extended attribution techniques to source code, motivated by applications in intellectual property protection and forensic analysis~\cite{b2,b3}. Code stylometry work highlights signals such as identifier naming conventions, indentation styles, whitespace use, and control-flow constructs as reliable markers of programmer identity. More recent large-scale studies further confirmed that programmers leave persistent stylistic traces across projects and programming languages~\cite{b3}. However, these approaches often assume human-authored code, and their robustness decreases under adversarial transformations such as reformatting, obfuscation, or automated style normalization.

\subsection{Detection of Machine-Generated Text}
The widespread adoption of large language models (LLMs) such as GPT-3 and ChatGPT has sparked intense research into detecting AI-generated natural language. Early detectors leveraged statistical irregularities such as token frequency shifts, perplexity measurements, and unusual burstiness patterns in generated text~\cite{b4}. Visualization tools like GLTR provided interpretable interfaces for spotting suspicious language. Subsequent work moved toward embedding-based classifiers that train neural or classical models on vector representations of text, offering stronger performance by capturing deep semantic and syntactic cues~\cite{b5,b6}. However, studies consistently report fragility under distribution shift: detectors trained on one model family may fail when applied to outputs from another generator. These lessons from text motivate similar inquiries in code, where generative models produce outputs that are both syntactically valid and semantically functional.

\subsection{Detection of Machine-Generated Code}
In contrast to natural language, systematic efforts to detect AI-generated code are still emerging. Early explorations investigated handcrafted metrics such as abstract syntax tree (AST) depth, cyclomatic complexity, control-flow structure, and identifier naming conventions to differentiate human from machine code~\cite{b7}. These studies suggested that machine-generated code often exhibits uniform stylistic patterns and reduced variability compared to human-authored code. Parallel to this, embedding-based methods have leveraged pretrained encoders such as CodeBERT~\cite{codebert, graphcb}, originally designed for code search, summarization, and translation tasks. By encoding source code into dense semantic representations, these models provide a natural foundation for detection. More recent analyses of tools like GitHub Copilot and ChatGPT outputs underscore both the potential of embeddings and the risks associated with over-reliance on black-box models, noting that while embeddings capture deeper semantics, they require significant computational resources and offer limited interpretability for stakeholders such as educators and policy makers~\cite{b11,b12}. 

At the same time, a new generation of specialized detectors has emerged. MageCode, for example, combines semantic features with code-specific metrics to achieve high detection accuracy~\cite{pham2024magecode}. CoDet-M4 extends this line of work by explicitly targeting generalization across multiple languages, domains, and generator families~\cite{azizov2025codetm4}. Zero-shot approaches based on large pretrained encoders~\cite{yang2023zeroshot} and stylistic pattern detectors such as DetectCodeGPT~\cite{shi2024detectcodegpt} further emphasize robustness under distribution shift. These efforts highlight the growing recognition that code-origin detection is not only an academic challenge but also a practical necessity for maintaining authenticity and accountability in educational and industrial contexts~\cite{hoq2024chatgpt}.

While recent methods such as MageCode~\cite{pham2024magecode}, CoDet-M4~\cite{azizov2025codetm4}, and DetectCodeGPT~\cite{shi2024detectcodegpt} push the frontier of machine-generated code detection by incorporating semantic features, multilingual robustness, or zero-shot capabilities, most of these approaches rely heavily on deep pretrained encoders and complex pipelines. These designs offer strong generalization but often come at the cost of high computational overhead, limited interpretability, and reduced accessibility for institutions with constrained resources. 
Our work addresses this gap by directly comparing lightweight, feature-based classifiers with embedding-based detectors on the same large-scale benchmark dataset. This design allows us to quantify trade-offs between efficiency, interpretability, and robustness, and to highlight when lightweight approaches may be sufficient and when embedding-based methods provide complementary advantages.


\section{Problem Description}

We formalize the task of machine-generated code detection as a supervised binary classification problem.  
Let $\mathcal{D} = \{(x_i, y_i)\}_{i=1}^N$ denote a dataset of $N$ code snippets, where each $x_i$ is a source code sample and $y_i \in \{0,1\}$ is the corresponding ground-truth label, with
\begin{equation}
y_i =
\begin{cases}
0, & \text{if $x_i$ is human-written}, \\
1, & \text{if $x_i$ is machine-generated},
\end{cases}
\label{eq:labels}
\end{equation}

Each code snippet $x_i$ is represented either through handcrafted features $f(x_i) \in \mathbb{R}^d$ or through dense embeddings $e(x_i) \in \mathbb{R}^h$ obtained from a pretrained model such as CodeBERT. The goal of the classifier is to learn a mapping
\begin{equation}
\hat{y}_i = g(\phi(x_i); \theta),
\label{eq:mapping}
\end{equation}
where $\phi(\cdot)$ is the representation function (features or embeddings), $\theta$ are the classifier parameters, and $\hat{y}_i \in [0,1]$ is the predicted probability that $x_i$ is machine-generated.

Training proceeds by minimizing the binary cross-entropy loss,
\begin{equation}
\mathcal{L}(\theta) = -\frac{1}{N}\sum_{i=1}^N \Big[ y_i \log \hat{y}_i + (1-y_i)\log(1-\hat{y}_i) \Big],
\label{eq:loss}
\end{equation}
where $\mathcal{L}(\theta)$ is the loss function.

At inference time, the classifier outputs $\hat{y}_i$, which is converted into a discrete prediction as
\begin{equation}
\tilde{y}_i =
\begin{cases}
1, & \text{if } \hat{y}_i \geq \tau, \\
0, & \text{otherwise},
\end{cases}
\label{eq:decision}
\end{equation}
where $\tau \in [0,1]$ is a decision threshold calibrated on validation data (e.g., to maximize the F1 score).  

The central challenge is to design $\phi(\cdot)$ and select $g(\cdot;\theta)$ such that the detector generalizes across code generators, languages, and unseen distributional shifts while balancing accuracy, interpretability, and computational cost.

\section{Methodology}

We adopt a comparative framework to detect whether a given code snippet is written by a human or generated by a machine. Our design emphasizes two complementary paradigms. The first is a \emph{feature-based} approach, in which we extract handcrafted metrics that capture surface-level, stylistic, and structural properties of code. These features have their roots in prior work on code stylometry and authorship attribution and offer interpretability and efficiency. The second is an \emph{embedding-based} approach, where pretrained language models for code are used to encode snippets into dense semantic vectors, enabling classifiers to exploit deeper contextual regularities. By training and evaluating both pipelines under a unified protocol, we systematically compare their strengths and limitations, with particular attention to interpretability, generalization, and computational cost.

\subsection{Data Source}

Our experiments rely on a recently introduced benchmark dataset by Orel et al. \cite{semeval} proposed in prior work on human versus machine code classification. Each instance consists of a raw source code snippet and an associated label: $y = 0$ for human-written and $y = 1$ for machine-generated. Metadata also includes the programming language and, for generated cases, the model that produced the snippet. The dataset covers primarily Python programs, while also including other languages that reflect the diversity of real-world code generation. For evaluation, we perform an 80/20 stratified train–validation split to preserve class balance and enable consistent threshold calibration. The test set of 100k samples is reserved exclusively for final reporting of results.

\subsection{Feature Extraction}

For the feature-based pipeline, each code snippet $x$ is represented by a feature vector
\begin{equation}
f(x) = \big[f_1(x), f_2(x), \dots, f_d(x)\big] \in \mathbb{R}^d,
\label{eq:features}
\end{equation}
where each component corresponds to a handcrafted metric. Features are grouped into three categories. 

\textbf{Surface statistics} include the number of lines, characters, average line length, blank-line and comment ratios, and the presence of docstrings, capturing structural regularities observed in human coding practices~\cite{b1}.  
\textbf{Identifier stylometry} focuses on identifiers and naming patterns, such as counts, average length, snake- vs.\ camel-case ratios, entropy of tokens, and uniqueness, which have been shown effective in distinguishing coding styles~\cite{b2,b3}.  
\textbf{Structural proxies} approximate higher-level program complexity, including estimated AST depth, counts of loops, conditionals, imports, comprehensions, and cyclomatic complexity. These metrics draw from classic software engineering measures such as McCabe’s complexity~\cite{mccabe1976complexity} and recent analyses of AI-generated code~\cite{b7}.  

Together, these features provide interpretable signals that reflect human habits and stylistic variation, as opposed to the more uniform patterns often produced by large language models.

\subsection{Classification Models}

On top of the handcrafted features, we evaluate a range of supervised classifiers spanning linear, nonlinear, and ensemble families. Linear and scaled models include Logistic Regression (LR), Support Vector Machines with RBF kernel (SVM–RBF),  and Multi-Layer Perceptrons (MLP). Ensemble methods include Random Forest (RF), Histogram-based Gradient Boosting (HGB), and ExtraTrees (ET). 

For example, Logistic Regression estimates the probability of machine generation as
\begin{equation}
\hat{p}(y{=}1 \mid x) = \sigma(\mathbf{w}^\top \mathbf{f}(x) + b),
\label{eq:logreg}
\end{equation}
where $\sigma(\cdot)$ is the sigmoid activation. In contrast, tree-based ensembles learn nonlinear boundaries through recursive partitioning of the feature space. All models are trained using binary cross-entropy loss,
\begin{equation}
\mathcal{L}_{\text{BCE}} = -\frac{1}{N}\sum_{i=1}^N \Big( y_i \log \hat{y}_i + (1-y_i)\log(1-\hat{y}_i)\Big).
\label{eq:bce}
\end{equation}

\subsection{Embedding-Based Detection}

In the embedding pipeline, each code snippet is encoded using the \texttt{microsoft/codebert-base} model. The encoder produces hidden states $H \in \mathbb{R}^{T \times h}$ for a sequence of length $T$, which are pooled into a fixed vector via attention masking:
\begin{equation}
\mathbf{e}(x) = \frac{\sum_{t=1}^{T} m_t H_t}{\sum_{t=1}^{T} m_t}, \quad
\tilde{\mathbf{e}}(x) = \frac{\mathbf{e}(x)}{\|\mathbf{e}(x)\|_2},
\label{eq:embedding}
\end{equation}
where $m_t$ is the attention mask. The normalized 768-dimensional embedding $\tilde{\mathbf{e}}(x)$ is then passed into the same set of classifiers as in the feature-based pipeline. To reduce training cost, encoder weights are frozen and only the downstream classifier is trained.

\subsection{Model Selection and Threshold Calibration}

For both pipelines, candidate classifiers are compared on the validation split. Model selection prioritizes Precision–Recall Area Under Curve (PR–AUC), with ROC–AUC as a secondary criterion. Since classifiers output probabilities, we calibrate a decision threshold $\tau^\ast$ to maximize the F1 score:
\begin{equation}
\tau^\ast = \arg\max_{\tau} \; \mathrm{F1}(\tau), \qquad 
\mathrm{F1} = \frac{2 \cdot P \cdot R}{P + R},
\label{eq:threshold}
\end{equation}
where $P$ and $R$ denote precision and recall at threshold $\tau$. This calibration is particularly important given the class imbalance in the test set.

\subsection{Evaluation Metrics}
Final evaluation is performed on the held-out test set. To provide a comprehensive assessment, we report multiple complementary metrics.  

We first include the Precision--Recall Area Under the Curve (PR--AUC) and the Receiver Operating Characteristic Area Under the Curve (ROC--AUC). Both are threshold-independent metrics that summarize model performance across all possible classification thresholds, but they emphasize different trade-offs. ROC--AUC measures the ability of a classifier to separate the two classes by plotting the true positive rate against the false positive rate. It is widely used, but can be overly optimistic when the data is imbalanced. PR--AUC, on the other hand, focuses on the balance between precision and recall, and is more informative when the positive class is rare or when false positives are particularly costly. Since our task involves distinguishing between human-written and machine-generated code under potential class imbalance, PR--AUC is a particularly relevant metric.
In addition, we report standerd classification evaluation metric \textbf{accuracy},  \textbf{F1 score},\textbf{precision} and \textbf{recall}.


\section{Results}

In this section, we present the results of our comparative study on detecting human-written versus machine-generated code using 500k training samples and 100k test samples. Our analysis is organized to address three guiding questions:  
\begin{itemize}
    \item How well do handcrafted feature-based models perform in-domain, and how do they generalize to unseen test data?  
    \item How do embedding-based models perform under the same experimental setup, and do they offer advantages in robustness?  
    \item What are the relative trade-offs between the two approaches in terms of accuracy, interpretability, and resilience to distribution shift?  
\end{itemize}

The following subsections report validation performance, test performance, and a direct comparison between the feature-based and embedding-based approaches.

\subsection{Validation Performance}
Table~\ref{tab:val_results} reports validation metrics across candidate models. For feature-based classifiers, ensemble methods such as Random Forest, Histogram Gradient Boosting, and ExtraTrees achieved the strongest results, with PR--AUC consistently above 0.994. For embedding-based classifiers, Logistic Regression outperformed more complex models, highlighting the near-linear separability of CodeBERT embeddings.

\begin{table}[b]
\caption{Validation performance of feature-based and embedding-based models.}
\label{tab:val_results}
\centering
\begin{tabular}{l|c|c}
\toprule
\textbf{Model (Features)} & \textbf{PR--AUC} & \textbf{ROC--AUC} \\
\midrule
Logistic Regression & 0.949 & 0.953 \\
Random Forest       & \textbf{0.995} & \textbf{0.995} \\
Histogram GB        & 0.994 & 0.994 \\
 SVM (RBF)           & 0.991 & 0.991 \\
 MLP                 & 0.993 & 0.993 \\
ExtraTrees          & 0.995 & 0.995 \\
\midrule
\textbf{Model (Embeddings)} & \textbf{PR--AUC} & \textbf{ROC--AUC} \\
\midrule
Logistic Regression & \textbf{0.994} & \textbf{0.993} \\
Random Forest       & 0.991 & 0.990 \\

Histogram GB        & 0.991 & 0.990 \\
SVM (RBF)           & 0.990 & 0.990 \\

MLP                 & 0.990 & 0.992 \\
ExtraTrees          & 0.989 & 0.989 \\
\bottomrule
\end{tabular}
\end{table}

The best-performing models on validation were Random Forest (feature-based) with F1 = 0.971 and Logistic Regression (embedding-based) with F1 = 0.965.

\subsection{Test Performance}
On the unseen test set (100k rows), both pipelines generalized strongly, but with subtle differences. The feature-based Random Forest achieved slightly higher F1 and recall, while the embedding-based Logistic Regression achieved marginally higher precision. Detailed metrics are presented in Table~\ref{tab:test_metrics}.

\begin{table}[b]
\caption{Test Performance of Best Models of Feature based approach and Embedding based approach (Unseen 100k data samples)}
\label{tab:test_metrics}
\centering
\resizebox{\columnwidth}{!}{
\begin{tabular}{l|c|c|c|c|c}
\toprule
\textbf{Approach} & \textbf{ROC--AUC} & \textbf{PR--AUC} & \textbf{F1} & \textbf{Accuracy} & \textbf{Precision}\\
\midrule
Features (RF) & \textbf{0.995} & \textbf{0.995} & \textbf{0.971} & \textbf{0.970}  & 0.961\\
Embeddings (LR) & 0.994 & 0.994 & 0.965 & 0.964  & \textbf{0.969}\\
\bottomrule
\end{tabular}
}
\end{table}

\begin{figure}[t]
    \centering
    \includegraphics[width=0.48\textwidth]{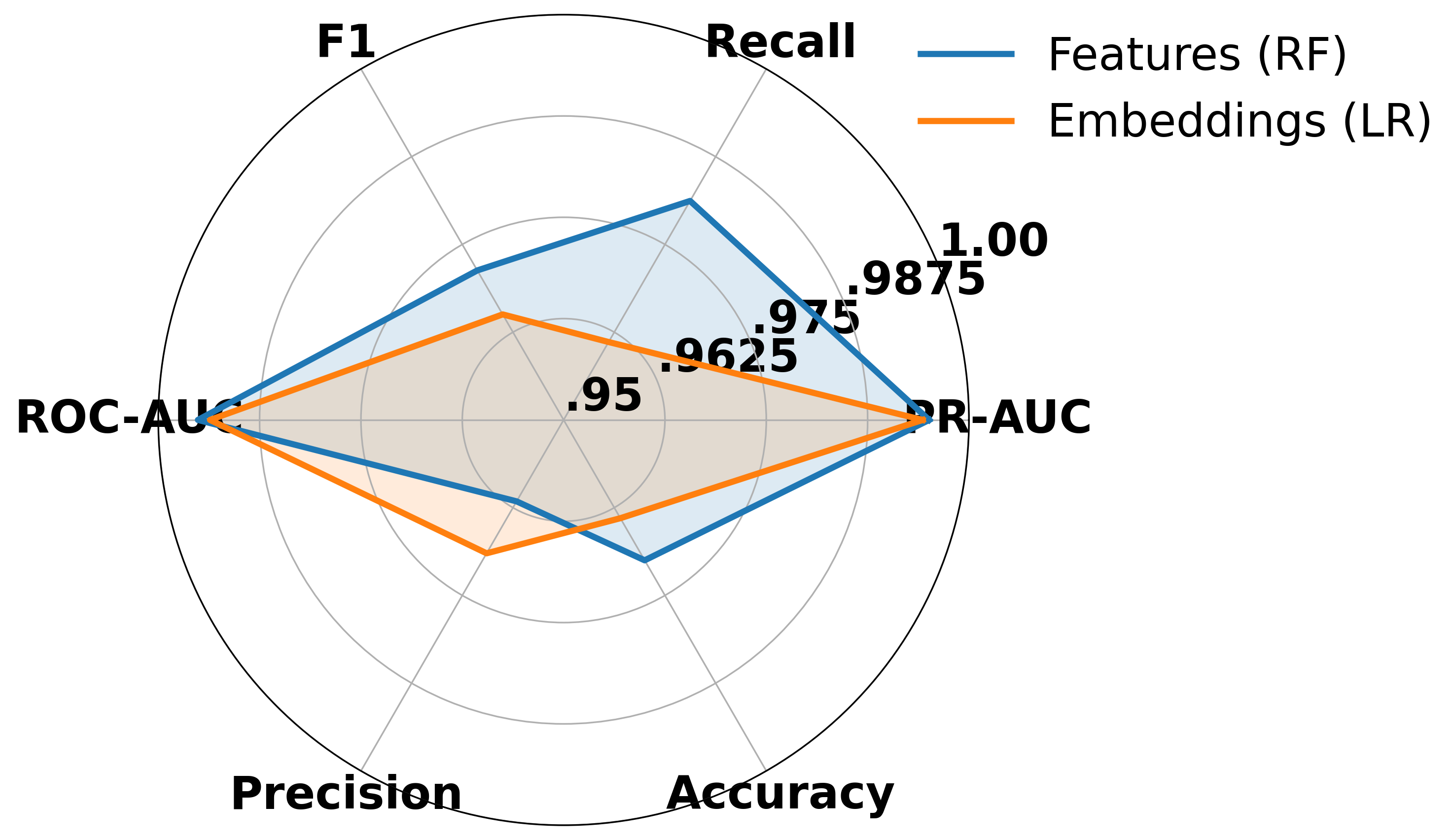}
   \caption{Radar plot comparing the best feature-based model (Random Forest) and embedding-based model (Logistic Regression) on the test set. PR--AUC and ROC--AUC are nearly identical, but the feature-based model achieves higher accuracy, F1 and recall, while the embedding-based model achieves slightly better precision. The plot is zoomed into the range 0.95--1.00 to better highlight subtle differences between the models.}

    \label{fig:radar_models}
\end{figure}

To highlight the trade-offs between the two approaches, Figure~\ref{fig:radar_models} shows a radar plot comparing ROC--AUC, PR--AUC, F1, precision, and recall for the two best-performing models. Both models achieve nearly identical ROC--AUC and PR--AUC, indicating comparable ranking ability. However, the feature-based model excels in F1 and recall, suggesting stronger coverage of AI-generated code, while the embedding-based model demonstrates slightly higher precision, reducing false positives.

\begin{figure}[t]
    \centering
    \includegraphics[width=0.48\textwidth]{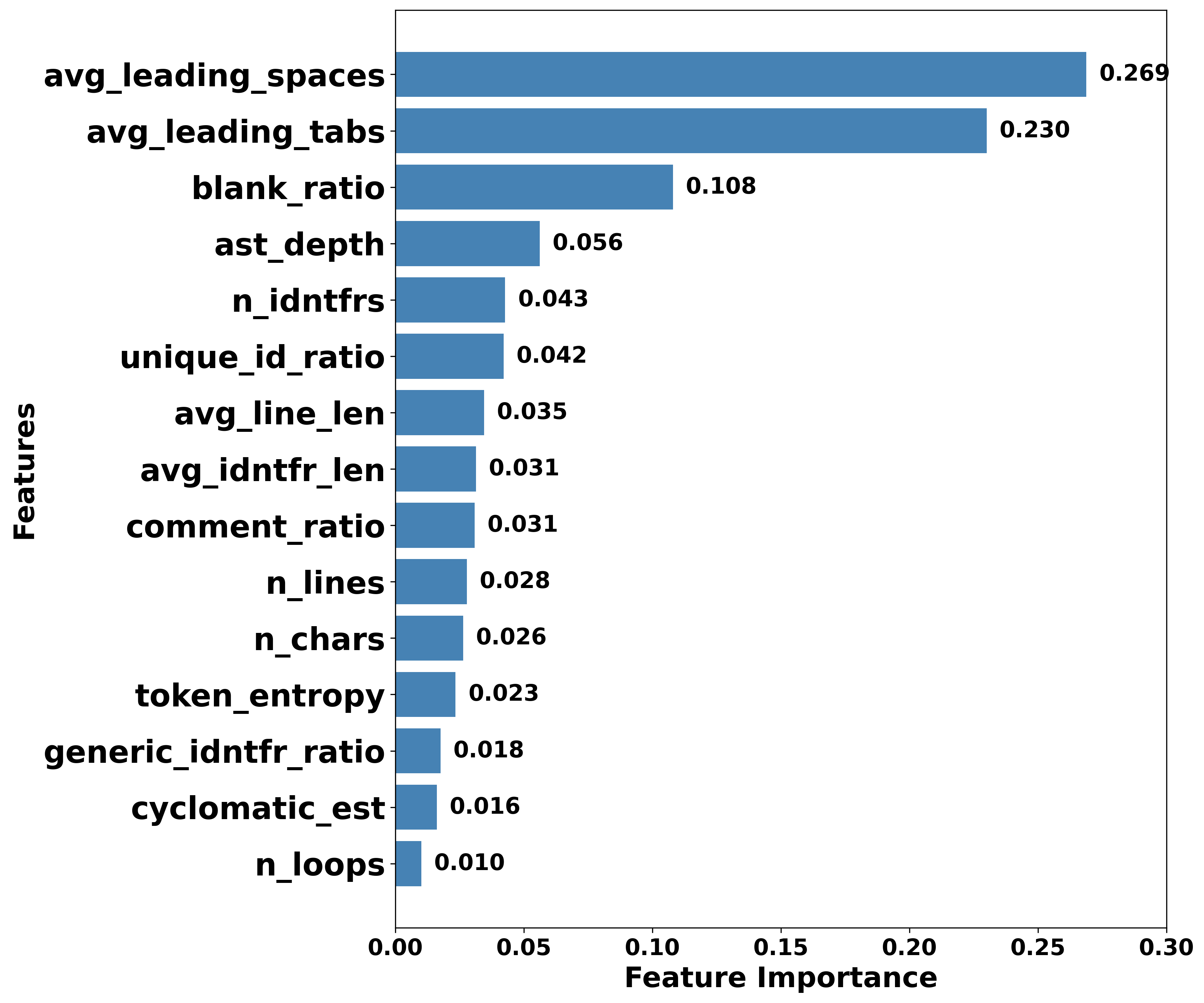}
    \caption{Feature importance ranking from the Random Forest classifier trained on handcrafted metrics. 
    The most discriminative features include \textbf{average leading spaces}, \textbf{average leading tabs}, 
    and \textbf{blank-line ratio}, followed by \textbf{AST depth}. 
    These results highlight that stylistic and structural cues such as indentation habits and whitespace 
    patterns carry strong signals for distinguishing human-written from machine-generated code.}
    \label{fig:feature_importance}
\end{figure}

For the feature-based Random Forest, the most discriminative features included average leading spaces, average leading tabs, blank-line ratio, and estimated AST depth. Figure~\ref{fig:feature_importance} shows the ranked importance values across all handcrafted features. The reported importance values are derived from the average reduction in impurity across decision nodes in the Random Forest ensemble, where features that consistently yield larger impurity decreases receive higher normalized scores.  The dominance of whitespace-related cues indicates that indentation and stylistic patterns strongly separate human-written from machine-generated code. At the same time, structural signals such as AST depth also contributed meaningfully, suggesting that both low-level formatting choices and higher-level code organization patterns play a role in distinguishing authorship. This analysis quantitatively confirms that whitespace usage, indentation habits, and structural depth are the strongest predictive signals for separating human and AI-generated code.


\section{Discussion}
The results presented in Section~V demonstrate that both feature-based and embedding-based approaches can effectively detect whether code is written by a human or generated by an AI system. However, a closer analysis reveals important trade-offs, strengths, and limitations that merit discussion.

First, the feature-based pipeline, particularly when combined with ensemble models such as Random Forest and Histogram Gradient Boosting, achieved the strongest overall performance in terms of F1-score and recall. This indicates that handcrafted features, capturing stylistic and structural patterns such as indentation depth, blank-line ratio, and abstract syntax tree (AST) depth, remain highly informative. The feature importance analysis (Figure~\ref{fig:feature_importance}) further confirmed that whitespace and indentation—often subconscious in human programmers but systematically produced by models—are key discriminative signals. These findings resonate with prior research in code stylometry and authorship attribution, where stylistic markers such as naming conventions, indentation, and whitespace patterns have been shown to serve as stable identifiers of authorship.

At the same time, embedding-based detectors using CodeBERT demonstrated competitive results, achieving slightly higher precision than the feature-based models. This suggests that semantic-level representations learned from large-scale pretraining can reduce false positives, thereby minimizing cases where human-written code is incorrectly flagged as AI-generated.  The radar plot comparison (Figure~\ref{fig:radar_models}) highlighted this trade-off clearly: feature-based models excel in recall and F1, whereas embedding-based models are more balanced, particularly in precision.

Another key observation lies in generalization. Both approaches achieved near-perfect validation metrics (ROC–AUC and PR–AUC above 0.99) on the training distribution. Yet, their behaviors on the held-out test set revealed differences: feature-based models, while still strong, showed greater sensitivity to distribution shifts compared to embeddings.  Embeddings, by contrast, appear more robust to such shifts, likely due to their ability to encode deeper semantic and syntactic information beyond superficial style.

Taken together, the findings show that feature-based and embedding-based methods are complementary rather than competing. Feature-based models offer interpretability, efficiency, and strong recall, making them effective for large-scale screening where coverage is critical. In contrast, embedding-based models achieve a better balance of precision, reducing false positives when reliability is essential. A hybrid or ensemble approach that leverages both can provide a practical trade-off between interpretability and robustness, ensuring resilience across code styles, programming languages, and unseen AI generators.

Beyond technical performance, these results highlight broader implications. As AI code generation grows in education and industry, detection tools will be vital for maintaining academic integrity, enabling fair evaluation, and supporting responsible AI adoption. The study demonstrates that interpretable, lightweight methods remain highly competitive and should not be overshadowed by more complex embedding models, while embedding-based approaches contribute essential robustness. Together, they point toward hybrid strategies as a promising direction for future research, which we further discuss in terms of implications, limitations, and ethics.

\section{Conclusion}
This work compared two complementary approaches for detecting machine-generated code: handcrafted feature engineering and transformer-based embeddings. Our results showed that feature-based models achieve the stronger recall and F1-scores, while embedding-based classifiers provide slightly higher precision, highlighting complementary strengths. These findings suggest that lightweight feature models remain competitive and interpretable, but embedding-based methods offer robustness under distribution shift. 

Overall, both approaches are valuable, and combining them may yield the most reliable detectors for academic integrity and responsible AI use. Future work should explore hybrid models, cross-language generalization, and adversarial robustness to strengthen real-world deployment. Beyond technical improvements, this line of research also informs policy and educational practices aimed at preserving fairness and accountability in the age of generative AI.
\bibliographystyle{IEEEtran}
\bibliography{refs}



\end{document}